\documentclass[12pt]{article}

\RequirePackage[colorlinks,citecolor=blue,urlcolor=blue]{hyperref}
\RequirePackage[OT1]{fontenc}
\RequirePackage{amsthm,amssymb,amsfonts,graphicx,subfigure,caption,bm,fullpage,wrapfig,setspace}

\usepackage{amsmath,amssymb,enumerate}
\usepackage{float}
\usepackage{algpseudocode,algorithm,algorithmicx}
\usepackage{mathrsfs}
\usepackage{multirow}
\usepackage{forest, adjustbox}
\usepackage{bbm}
\usepackage{lscape}
\usepackage{booktabs, tabularx}
\usetikzlibrary{calc}
\usepackage{relsize}

 \usepackage{accents}
 
\tikzset{fontscale/.style = {font=\relsize{#1}}
    }
\numberwithin{equation}{section}
\theoremstyle{plain}
                          \newtheorem{thm}{Theorem}
                          \newtheorem{lem}{Lemma}

\theoremstyle{remark}         
\theoremstyle{definition} 

\newcommand\bz{\bm{z}}
\newcommand\bZ{\bm{Z}}

\usepackage{tikz, amsmath, bbm, bm}
\usetikzlibrary{trees}
\usetikzlibrary{calc}

\tikzset{
  treenode/.style = {shape=rectangle, rounded corners,
                     align=right,
                     top color=white, bottom color=white},
  root/.style     = {treenode, bottom color=white},
  unobs/.style      = {shape=rectangle, rounded corners, text width = 4em,
                     align=center, color =gray,
                     top color=white, bottom color=white},
  dummy/.style    = {circle,draw}
}

\tikzstyle{level 1}=[level distance=4.5cm, sibling distance=3.5cm]
\tikzstyle{level 2}=[level distance=4.5cm, sibling distance=1cm]

\tikzstyle{bag} = [treenode, text width=4em, text centered]
\tikzstyle{end} = [circle, minimum width=0pt,  inner sep=0pt]

\newcommand{\RN}[1]{%
  \textup{\uppercase\expandafter{\romannumeral#1}}%
}

\newcommand\bi{\begin{itemize}}
\newcommand\ei{\end{itemize}}

\newcommand{\obs}{{\rm{obs}}}

\newcommand\independent{\protect\mathpalette{\protect\independenT}{\perp}}
\def\independenT#1#2{\mathrel{\rlap{$#1#2$}\mkern2mu{#1#2}}}

 \allowdisplaybreaks

\newcommand{\beginsupplement}{%
        \setcounter{table}{0}
        \renewcommand{\thetable}{S\arabic{table}}%
        \setcounter{figure}{0}
        \renewcommand{\thefigure}{S\arabic{figure}}%
        \renewcommand{\theequation}{S\arabic{equation}}
}

\theoremstyle{definition} \newtheorem{definition}{Definition}
                          
                          \newtheorem{assumption}{Assumption}

\def\woMR#1{\w@MR#1MR#1MR\relax}%
\def\w@MR#1MR#2MR#3\relax{#2}

\def\@MR#1 #2\relax#3{%
 \href{http://www.ams.org/mathscinet-getitem?mr=#1}%
 {\MRfixed{#3}}}%

\def\MRfixed{MR\woMR}%

\usepackage{natbib}
\usepackage{varioref}		
\labelformat{section}{Section~#1}
\labelformat{figure}{Figure~#1}
\labelformat{table}{Table~#1}	

\title{Quantifying the Value of Iterative Experimentation}
\author{Jialiang Mao \\ \textit{LinkedIn Corporation} \and Iavor Bojinov \\ \textit{Harvard Business School}}
\date{\today}
\begin{document}
\maketitle

 \vspace{-3mm}
\begin{abstract}
\doublespacing  
Over the past decade, most technology companies and a growing number of conventional firms have adopted online experimentation (or A/B testing) into their product development process. Initially, A/B testing was deployed as a static procedure in which an experiment was conducted by randomly splitting half of the users to see the control—the standard offering—and the other half the treatment—the new version. The results were then used to augment decision-making around which version to release widely. More recently, as experimentation has matured, firms have developed a more dynamic approach to experimentation in which a new version (the treatment) is gradually released to a growing number of units through a sequence of randomized experiments, known as iterations. In this paper, we develop a theoretical framework to quantify the value brought on by such dynamic or iterative experimentation. We apply our framework to seven months of LinkedIn experiments and show that iterative experimentation led to an additional 20\% improvement in one of the firm's primary metrics. 
\end{abstract}

\doublespacing

\clearpage

\section{Introduction}
\label{sec:intro}
Over the past decade, firms, particularly in the technology sector, have adopted randomized experiments (often referred to as A/B tests) to guide product development and innovation \citep{thomke2020experimentation}. Companies including Airbnb \citep{lee2018winner}, Google \citep{tang2010overlapping}, LinkedIn \citep{xu2015infrastructure}, Microsoft \citep{kohavi2013online} and Yelp \citep{BojinovCaseYelp2020}  all developed in-house experimentation platforms to enable large-scale experimentation. Commercial A/B testing platforms such as Optimizely and Split have also recently emerged to serve experimentation as a product to enable traditional firms as well as start-ups to gain access to this valuable tool \citep{johari2017peeking}.

The operational benefits have primarily fueled the rapid uptick in the integration of experimentation. We can categorize these into three pillars: reduced risk of product launches and changes, fast product feedback, and accurate measurement of the effect of innovations \citep{kohavi2009online,xia2019safe,xu2018sqr}. There is an abundance of anecdotal evidence for each pillar that illustrates the value to a firm \citep{kohavi2017surprising,koning2019experimentation}. Most of this work focuses on A/B test as a two-step procedure: an experiment is conducted, and a decision is made based on the results \citep{kohavi2013online}. However, as firms reach a high level of experimentation maturity and integrate experimentation into most product changes, experimentation becomes an iterative process where a feature change is gradually released and is quickly changed based on the user feedback; this process is known as a phased release \citep{xu2018sqr}. 

A phased release cycle is divided into three steps; usually, firms decide to move from one step to the next by examining the estimated impact of a change on multiple business metrics \citep{xu2018sqr}. First, the change is introduced to a small random sample of users; this stage is used to detect degradations. Second, we increase the proportion of exposed users to 50\%; at this point, we have maximum power to detect a change in business metrics. Third, we choose to either make the change available to all users or roll back to the original version. In practice, we found that most teams use the first step to learn how to improve their product and fix bugs. The second step then measures the overall impact of the innovation and, in almost all cases, is precisely the final released product. Changes can be made to improve the product between the two steps, suggesting that we can quantify the value of iterative experimentation by comparing the estimated effect in the first step to the estimate obtained from the second step. This paper formalizes this intuition and develops a framework for quantifying the value of transitioning from a two-step experimentation process to an iterative approach. Our framework inherently captures aspects of the value of three pillars but does not capture every potential benefit of A/B testing. When applied to seven months of experiments at LinkedIn,  our results suggest that iterative experimentation, or phased release, leads to an additional 20\% improvement in underlying metrics. 

The key to our formulation is a framework that describes the inherently temporal nature of A/B testing. Instead of focusing on one iteration at a time, we consider multiple iterations of an experiment in a holistic way, which allows us to combine information from different iterations and provides new perspectives to analyze the experiment results. In particular, we define the value of an iteration experiment as a special causal estimand that includes potential outcomes from multiple iterations to quantify the improvement we gain during the experimentation period. Moreover, the general framework we build makes it easy to elucidate the assumptions involved in identifying and estimating our causal estimand, especially those related to the dynamics of the potential outcomes, which are usually made in practice in an ambiguous and implicit way.

Interestingly, by combining the estimated value of a single experiment across all experiments, we can obtain an overall estimate of the value of iterative experimentation. In particular, the value of multiple experiments in a fixed time window can be aggregated to estimate the overall value delivered by the transition to iterative experimentation. This estimate can be considered as a lower bound of the value provided by an experimentation platform that enables rapid iteration. Practically, it also allows us to define several metrics that track the performance of the experimentation platform. First, such aggregation can be done at the company level to provide metrics that directly evaluate the platform's overall performance. Unlike qualitative feedbacks or existing metrics used to assess the platform, such as the experiment capacity and average experiment velocity, these value-based metrics directly measure the platform's performance in terms of key business metrics. Second, the iterative value of experiments from the same team can be aggregated to quantify the learning and innovation level of that team. Such metrics can provide insights into the dynamic of the company's product space and give feedback to the platform to further boost innovation.

Our proposed approach to quantify the value of iterative experiments has a ``Bootstrapping" nature: everything needed to compute the value of an experiment is already contained in the data collected by a typical experimentation platform. To compute the value, the practitioner needs to combine data from different iterations when computing the confidence intervals or $p$-values. In most mature platforms, such data are usually tracked and stored in offline databases in an easy-to-query manner \citep{gupta2018anatomy}.  Thus there is essentially no engineering effort required to implement our proposal. From an information collection perspective, the introduced metrics to quantify the value of an experimentation platform synthesizes information from multiple randomized controlled experiments. It thus has a flavor of meta-analysis of finished experiments and sits on top of the ``hierarchy of evidence" \citep{kohavi2020trustworthy}. 

This paper is organized as follows. \ref{sec:setup} introduces the experimentation design and defines the potential outcomes. In \ref{sec:single}, we define and estimate the value of a single iterative experiment under two slightly different designs. The aggregated value of an experimentation platform is then introduced in \ref{sec:platform}. In \ref{sec:application}, we present our analysis on the value of LinkedIn's experimentation platform in the first half of 2019. \ref{sec:conclude} concludes.


\section{Treatments and potential outcomes}
\label{sec:setup}

\subsection{Formalizing the phased release with changing features}

We first focus on a single feature being launched through a phased release and denote the outcome of interest as $Y$. Suppose that the target population is fixed and contains $n$ units denoted by $i\in [N] = 1,2,\ldots, n$. This fixed population perspective is attractive in our setting as the units often comprise the entire population of interest; see \cite{abadie_2020sampling} for a review. A phased release is typically implemented as a sequence of iterative experiments, each of which is often called a ramp \citep{xu2018sqr}. In each iteration, the treatment is assigned to a randomly selected group of units. The treatment group starts small and expands across different iterations before it reaches the entire target population when the feature is fully launched. Companies in the tech industry usually have recommendations on their ramping plans \citep{apple_2021, google_2021}. For example, LinkedIn limits the set of potential ramping probabilities of its internal experiments to $\{1\%, 5\%, 10\%, 25\%, 50\%\}$ for standardization and communication purpose. 

The first few iterations before the $50\%$ ramp correspond to the \textit{risk mitigation stage}. Ramps in this stage serve as test runs of the new feature and collect timely feedback to help make product launch decisions and guide future product development. In addition, by exposing the change only to a small group of users, the risk of launching a poor feature or introducing unexpected bugs to the system is significantly reduced. Ramps at $50\%$ correspond to the \textit{effect estimation stage} of a phased release and are usually referred to as the \textit{``max power iterations" (MP)}. As the name suggests, it is straightforward to show\footnote{An experiment with an even split between the treatment and control group minimizes the variance of the $t$-statistic used to analyze the results.} that this ramp offers maximum power to detect differences between the treatment and control and provides the most accurate estimate of the treatment effect.

Practitioners assume that the feature remains the same throughout the release. However, this assumption is often violated as developers use the feedback in earlier iteration to improve the feature. Technically, if a feature is updated, it should go through an entirely new phased release cycle. However, many of these changes are minor and do not introduce additional risks. As the first few iterations in the cycle are mainly for risk mitigation, starting over from the first iteration is undesirable whenever the changes are minor as it will introduce significant opportunity costs and slow down the pace of innovation. In practice, a feature after minor changes is often considered the same feature as long as these changes do not modify the product at a conceptual level. A new phase release cycle is only required when the changes are significant enough to alter the essence of the product.

For example, recently, at LinkedIn, a team released a new brand advertising product with a new auto bidding strategy to improve advertiser return on investment by maximizing the number of unique LinkedIn members reached by an advertising campaign. There was an overall improvement in core metrics in the first iteration, like the number of unique members reached and the click-through rate. However, further analysis suggested the change caused a drop in the budget utilization of campaigns with a small target audience, decreasing the revenue from this segment. The developers used the feedback and identified the problem:  a global parameter in the new algorithm that determined the auction behavior of all campaigns. The team then came up with a solution to make the parameter adaptive and more campaign-specific. Follow-up iterations showed that the adaptive parameter fixed the identified issue and was eventually launched. Although modifications were made to the new bidding strategy across iterations, the concept of this product did not change. 

Here, we formally consider such modifications to the treatment with the following restrictions. Firstly, changes to the treatment must not alter the treatment at the conceptual level and cannot introduce additional risks. Secondly, all changes must be made in the risk mitigation stage, and no changes are allowed after the $50\%$ ramps. Thirdly, the final version of the treatment has to be tested with at least one $50\%$ ramp to estimate its effect accurately. With these requirements, we focus on two iterations that provide sufficient information to quantify the value of an iterative experiment: the \textit{largest unchanged iteration (LU)} and the \textit{last most powerful iteration (LMP)}. Specifically, an unchanged iteration is an iteration in which the treatment is in its original form without any modification. LU is the unchanged iteration with the largest treatment ramping percentage. Intuitively, LU gives the most accurate estimate of the effect of the original treatment, while LMP estimates the effect of the final version of the treatment after incorporating learnings from previous iterations. In practice, LU can range from the first iteration to MP. If changes are not clearly documented, it is always safe to pick the first iteration as LU. 

The new feature being tested in an phased release can have different variants. For example, we might test different color schemes in an user interface change, different ranking algorithms in a search engine, or different values of the bidding floor price in an ads serving system. For illustration, we first assume that the feature only has a single variant or version and denote it as $v$. Let $t=1,2$ denote the LU and the LMP respectively. We write the potentially different versions of $v$ in iteration $t$ as $v_t$ for $t=1,2$. For unit $i$, let $Z_{i,t} \in \{ v_1, v_2, c \}$ be the product treatment she receives in iteration $t$ and let $\bZ_i=(Z_{i,1}, Z_{i,2})$ be her \textit{treatment path} \citep{bojinov2019time, bojinov2020panel}. We also use $\bZ_{1:N}$ to represent the $N\times 2$ matrix of the treatment paths of all units. We assume that the treatment assignment follows a two-iteration stepped-wedge design \citep{brown2006stepped} in which the units who receive the treatment in the first iteration will stay in the treatment group in the second iteration. Formally, we have
\begin{definition}[Two-iteration stepped-wedge design]
\label{def:design}
Let $\mathcal{H} = \{ \bZ_{1:N}: \bZ_i \in \mathcal{V}= \{ (c,c), (c,v_2),$ $ (v_1,v_2) \}, i\in [N] \}$, a two-iteration stepped-wedge design $\eta = \eta(\bZ)$ is a probability distribution on $\mathcal{H}$.  
\end{definition}

This stepped-wedge design can be implemented by introducing restrictions on the sequential assignment at the two iterations. For example, we can implement a completely randomized design or a Bernoulli design at each iteration as long as we respect the ``no going back" constraint. In this work, we focus on the completely randomized design and let $0 < p_1 \leq p_2 = 0.5$ be the ramping probabilities at the two iterations. In the first iteration, we randomly assign $p_1N$ units to the treatment group. In the second iteration, we keep these units in the treatment group and randomly assign an additional $(p_2 - p_1) N$ units from the previous control group to the treatment group. Most technology companies adopt the stepped-wedge design in phased releases to ensure a consistent product experience. It is also a common setup in medical and social science research where it is sometime referred to as the staggered adoption design or the event study design \citep{brown2006stepped, callaway2020difference, sun2020estimating, athey2021design}.


\subsection{Potential outcomes}
\label{sec:po}
We adopt the Neyman-Rubin potential outcome framework for causal inference \citep{splawa1990application, rubin1974estimating} and treat the potential outcomes as fixed. Specifically, let $Y_{i,t}(\bZ_{1:N})$ be the potential outcome of unit $i$ at iteration $t$ under treatment path $\bZ_{1:N}$. In practice, each iteration of a phased release usually runs several days and there is no guarantee that different iterations will last the same duration. For example, the first iteration typically runs shorter than the second iteration. The majority of experiments at LinkedIn finish both iterations within a week \citep{xu2018sqr}. Therefore, there is a time component underlying the potential outcomes. In this paper, we interpret the potential outcome as the daily average of metric $Y$ measured over each iteration. Other interpretations such as the aggregated measure within each iteration can also be applied. 

We make the following assumptions, introduced in \cite{bojinov2019time}, to limit the dependence of the potential outcomes on assignment paths:
\begin{assumption}[Non-anticipation]
\label{ass:nonanticipation}
    The potential outcomes are non-anticipating if for all $i\in [N]$, $Y_{i,1}(\bz_{1:N, 1}, \bz_{1:N, 2} ) = Y_{i,1}(\bz_{1:N, 1}, \tilde\bz_{1:N,2})$ for all  $\bz_{1:N, 1} \in \{c, v_1 \}^N$, and $\bz_{1:N, 2},\tilde\bz_{1:N, 2} \in \{c, v_2\}^N$.
\end{assumption}
\begin{assumption}[No-interference]
\label{ass:nointerference}
    The potential outcomes satisfy no-interference if for all $i\in[N]$ and $t=1,2$  $Y_{i,t}( \bz_{1:(i-1)}, \bz_i, \bz_{(i+1):N}) = Y_{i,t}(\tilde\bz_{1:(i-1)}, \bz_i, \tilde\bz_{(i+1):N})$ for all $\bz_{1:(i-1)},\tilde\bz_{1:(i-1)}\in\mathcal{V}^{i-1}$ and $\bz_{(i+1):N},\tilde\bz_{(i+1):N}\in\mathcal{V}^{N-i}$.
\end{assumption}
With these assumptions, we can write the potential outcomes of unit $i$ as $Y_{i, 1}(Z_{i, 1})$ and $Y_{i, 2}(Z_{i, 1}, Z_{i, 2})$. Illustrations of the potential outcomes under the stepped-wedge design given $p_1 < p_2$ or $p_1 = p_2$ are shown in \ref{fig:po} and \ref{fig:po2}. Note that although practically $v_2$ is only available after the first iteration, we can still define $Y_{i,1}( v_2)$ theoretically. These potential outcomes capture the oracle scenario where the experimenter knows the best solution in the first place.


\section{Quantifying the operational value of iterative experimentation}
\label{sec:single}

\subsection{A single experiment}

There are three cases when an iterative experiment creates direct quantifiable value:
\begin{enumerate}
    \item The experiment iterations provide timely feedback and point the direction for further feature improvements. The value of the experiment can be quantified as the improvement in the treatment across the different iterations.
    \item When the feature has a negative impact and is de-ramped after the mitigation stage (i.e., reverted back to the control), the experiment helps prevent the exposure of bad features directly to a larger group of users. The value of these types of experiments can be estimated as the prevented loss.
    \item When the feature has multiple versions, the experiment can help identify the best possible version, avoiding random guesses on which version to launch and the cost of trial and error. The value of these types of experiments can be thought of as the difference between the best and worse performing versions. 
\end{enumerate}
In this section, we formally define these values and provide consistent estimators of them. 

We first introduce the value from feature improvement. Within each iteration ($t=1$ or $2$), the direct comparison of any two potential outcomes of the same unit has a causal interpretation. For example, $Y_{i,1}(v_1) - Y_{i,1}(c)$ is the effect of the treatment on unit $i$ in the first iteration and $Y_{i,2}(c, v_2) - Y_{i,2}(c, c)$ is the effect of the updated treatment on unit $i$ in the second iteration. Intuitively, the experiment provides value if $v_2$ is better than $v_1$. Below we define the improvement due to iterative experimentation. 
\begin{definition}[Value of iterative experimentation]
\label{def:voe}
We define the \textit{value of iterative experimentation (VOIE)} on unit $i$ as $\tau_{i, 1} =  Y_{i, 1}(v_2) -  Y_{i, 1}(v_1)$. We also define the population average VOIE as 
\begin{equation}
\begin{aligned}
\tau_{1} =\frac{1}{N}\sum\limits^{N}_{i=1}\tau_{i,1}= \frac{1}{N}\sum\limits^{N}_{i=1}[Y_{i, 1}(v_2) -  Y_{i, 1}(v_1)].
\end{aligned}
\label{eq:voie}
\end{equation}
\end{definition}
Similar to the potential outcomes, the interpretation of $\tau_{i,1}$ and $\tau_1$ also has to take into account the underlying time component. Firstly, the ``1" in the subscript indicates that these estimands capture the value of the experiment at the calendar time of the first iteration. If the experiment runs at a different time (i.e., in a different week), the value might change without further assumptions on the dynamics of the potential outcomes. Secondly, $\tau_{i,1}$ and $\tau_1$ represent the daily average value of the experimentation as inherited from the definition of the potential outcomes. The interpretation of VOIE can change if the experiment runs for a different duration (i.e., run for two weeks instead of one). 

In classical causal inference problems with a single treatment, one of the two potential outcomes of each unit is missing and we have to deal with the ``fundamental problem of causal inference" \citep{holland1986statistics}. To estimate the population level causal estimands, identifiability assumptions such as the strong ignorability assumption \citep{rosenbaum1983central} are often made. Our estimate of the population level VOIE relies on the following identifiability assumption:
\begin{assumption}[Time-invariant control effect]
\label{ass:constant}
   The individual-level treatment effects are time-invariant if $Y_{i,2}(c, v) - Y_{i,2}(c, c) = Y_{i, 1}(v) - Y_{i,1}(c)$ for $i\in [N], v\in \{v_1,v_2\}$.
 \end{assumption}
 
The time-invariant control effect assumption states that the effect of the treatment does not change with the time it is adopted. This assumption is justifiable in most cases as units were all assigned to control prior to the treatment release, therefore, an additional control exposure should have no bearing on the causal effect. Under this assumption, we can write $\tau_{i,1}$ as
\begin{equation}
\begin{aligned}
    \tau_{i,1} &= Y_{i,1}(v_2) - Y_{i,1}(v_1)  \\
     &=[Y_{i,1}(v_2) - Y_{i,1}(c)] - [Y_{i,1}(v_1) - Y_{i,1}(c)]  \\
     &= [Y_{i,2}(c, v_2) - Y_{i,2}(c, c)] - [Y_{i,1}(v_1) - Y_{i,1}(c)]
\end{aligned}
\end{equation}

The definition of VOIE could also capture other types of values of iterative experimentations. When the feature has a significantly negative effect and is de-ramped after the first iteration, the value of the experiment is the avoided potential loss. In this case, there is no second iteration in the experiment, and with a slight abuse of notation, we write ``$v_2 = c$". The VOIE in this case is 
\begin{equation}
\begin{aligned}
\tau_1^\prime = \frac{1}{N}\sum\limits^{N}_{i=1}[Y_{i, 1}(v_2) -  Y_{i, 1}(v_1)] =  - \frac{1}{N}\sum\limits^{N}_{i=1}\{[Y_{i, 1}(v_1) -  Y_{i, 1}(c) ] \},
\end{aligned}
\label{eq:multiple_variant}
\end{equation}
which is the opposite of the treatment effect in the first iteration. 

We need to slightly modify our notations to account for the case when the feature has multiple variants, and the practitioner needs to choose the best one to launch. Specifically, let $v^{(1)}, v^{(2)}, \ldots, v^{(m)}$ be the $m$ treatment variants under consideration. Suppose that the $m$ variants are tested on a randomly selected group of $p_1 N$ units in the first iteration. In this paper, we consider a completely randomized design with group size $(p_{11}, p_{12}, p_{1m}, 1-p_1)\cdot N$ respectively, where $\sum^{m}_{j=1}p_{1j}= p_1$. The ramping probabilities reflect the experimenter's prior knowledge and preference of different variants. Given feedback from the first iteration, one of the $m$ variants is picked to be further tested in the second iteration. We denote this winning variant as $v_2$. The potential outcomes defined in Section~\ref{sec:po} can be easily generalized to the case of multiple variants. Under assumption \ref{ass:constant}, we can define and simplify the VOIE in this case as
\begin{equation*}
\begin{aligned}
\tau_1^{*} &= \frac{1}{N}\sum\limits^{N}_{i=1}\left[Y_{i, 1}( v_2) -  \sum\limits^{m}_{j=1}\frac{p_{1j}}{p_1}Y_{i, 1}( v_1)  \right] \\
& =  \frac{1}{N}\sum\limits^{N}_{i=1}\left\{ [Y_{i, 2}( c, v_2) - Y_{i,2}(c,c)]-  [\sum\limits^{m}_{j=1}\frac{p_{1j}}{p_1} Y_{i, 1}( v_1) - Y_{i,1}(c)] \right\}.
\end{aligned}
\end{equation*}

Although $\tau_1$, $\tau_1^\prime$ and $\tau_1^*$ are all defined and estimated in a similar manner, they capture different aspects of the value of iterative experiments: $\tau_1$ estimates the value from guiding product development; $\tau_1^\prime$ captures the value from risk mitigation; and $\tau_1^*$ represents the value of reducing uncertainties. Both $\tau_1$ and $\tau_1^*$ can be viewed as capturing the value of learning and they can be combined. For example, in $\tau_1^*$, $v_2$ could be an improved version of the best variant in the first iteration. In the following sections, we provide estimates of the VOIE under two similar but slightly different designs. We will focus on $\tau_1$ as inference for $\tau_1^\prime$ and $\tau_1^*$ can be done in a similar manner.


\subsection{Estimating VOIE with progressive iterations}
\label{sec:single_1}

\begin{figure*}
\vspace{0.1cm}
\begin{center}
\tikzstyle{level 1}=[level distance=3.5cm, sibling distance=3.3cm]
\tikzstyle{level 2}=[level distance=3.5cm, sibling distance=1.6cm]
 \begin{tikzpicture}[->, grow=right, sloped]
\node[root] {Start}
    child {
        node[bag] { $Y_{i,1}(c)$}        
            child {
                node[bag] {$Y_{i,2}(c,c)$}
                edge from parent[black, line width=0.1mm]
                node[above] {}
                node[below]  {$$}
            }
            child {
                node[bag] {$Y_{i,2}( c,v_2)$}
                edge from parent[red, line width=0.5mm]
                node[above] {$$}
                node[below]  {}
            }
            edge from parent[red, line width=0.5mm]
            node[above] {$$}
            node[below]  {$$}
    }
    child {
        node[bag] { $Y_{i,1}(v_1)$}        
            child[dashed, gray] {
                node[unobs] {$Y_{i,2}( v_1,c)$}
                edge from parent
                node[above] {}
                node[below]  {}
            }
            child {
                node[bag] {$Y_{i,2}(v_1, v_2)$}
                edge from parent
                node[above] {$$}
                node[below]  {$$}
            }
            edge from parent 
            node[above] {$$}
            node[below]  {$$}
    }
;
\end{tikzpicture}
\end{center}
\caption{Potential outcomes for unit $i$ under the two-iteration stepped-wedge design when $p_1 < p_2$. The gray node represents the potential outcome when the feature is de-ramped after the first iteration. The red path shows one possible realization of the potential outcomes.}
\label{fig:po}
\end{figure*}
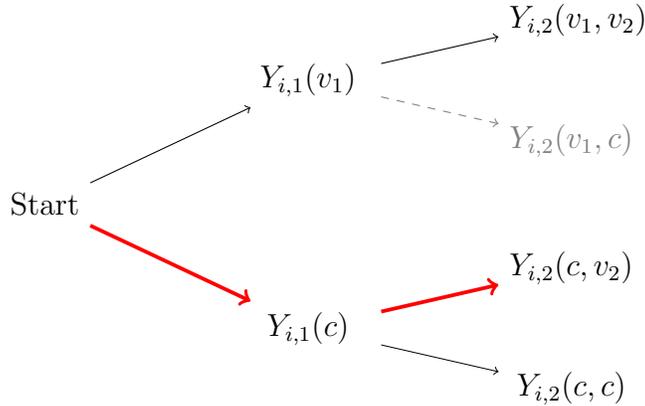

In practice, it is common that learning starts in the risk mitigation stage before the $50\%$ ramp. Minor changes to the feature like a bug fix in the code can be made right after the first iteration. In this case, the last unchanged iteration (LU) corresponds to a ramp with $p_1 < p_2 = 0.5$. An illustration of potential outcomes under this design is given in \ref{fig:po}. 

Under Assumption~\ref{ass:constant}, the population level VOIE $\tau_1$ in (\ref{eq:voie}) can be written as
\begin{equation}
\begin{aligned}
\tau_1 & =  \frac{1}{N}\sum\limits^{N}_{i=1}\{ Y_{i, 2}(c, v_2) -  Y_{i, 1}(v_1) -  \Delta_i ] \} ,
\end{aligned}
\label{eq:voie_ass}
\end{equation}
where $\Delta_i = Y_{i, 2}(c,c) -  Y_{i, 1}(c)$. 
To estimate $\tau_1$, we consider the following plug-in estimator 
\begin{equation}
\begin{aligned}
\hat\tau_1 &= \frac{1}{N_{c,v_2}}\sum\limits^{N}_{i=1}Y_{i,2}^{\obs} \mathbbm{1}(\bZ_i = (c, v_2)) - \frac{1}{N_{v_1}}\sum\limits^{N}_{i=1}Y_{i,1}^{\obs} \mathbbm{1}(Z_{i,1} = v_1) \\
&\quad  - \frac{1}{N_{c,c}} \sum\limits^{N}_{i=1}[Y_{i,2}^{\obs} \mathbbm{1}(\bZ_i = (c, c)) - Y_{i,1}^{\obs} \mathbbm{1}(\bZ_i = (c, c))],
\end{aligned}
\label{eq:plugin}
\end{equation}
where $N_{c,v_2} = \sum^{N}_{i=1} \mathbbm{1}(\bZ_i = (c, v_2))$ is the number of units receiving treatment path $(c, v_2)$, i.e., those assigned to the control group in the first iteration and to the treatment group in the second iteration; $N_{v_1}$ and $N_{c,c}$ are defined similarly. $Y_{i,1}^{\obs}$ and $Y_{i,2}^{\obs}$ are the observed outcomes of unit $i$ in the first and second iteration respectively. Let $\bar Y_{2}(c, v_2) = \sum^{N}_{i=1}Y_{i,2}(c, v_2)/N$, $\bar Y_{1}(v_1) = \sum^{N}_{i=1}Y_{i,1}(v_1)/N$, $\bar Y_{2}(c, c) = \sum^{N}_{i=1}Y_{i,2}(c, c)/N$ and $\bar Y_{1}(c) = \sum^{N}_{i=1}Y_{i,1}(c)/N$ be the population mean of potential outcomes under specific treatment paths. 
\begin{thm}
\label{thm:thm_1}
Under the two-iteration stepped-wedge design $\eta$ with $p_1 < p_2 $, the plug-in estimator in (\ref{eq:plugin}) satisfies 
\begin{equation*}
\begin{aligned}
\mathbbm{E}[\hat\tau_1] = \tau_1, \quad \mathbbm{V}[\hat\tau_1] = \frac{1}{N_{c,v_2}}S_{c,v_2}^2 +\frac{1}{N_{v_1}}S_{v_1}^2  +\frac{1}{N_{c,c}}S_{\Delta}^2 - \frac{1}{N}S_{\tau}^2,
\end{aligned}
\end{equation*}
where 
\begin{equation*}
\begin{aligned}
S^2_{c,v_2} & = \frac{1}{N-1}\sum\limits^{N}_{i=1}[Y_{i,2}(c,v_2)- \bar Y_{2}(c,v_2) ]^2, \quad S^2_{v_1} = \frac{1}{N-1}\sum\limits^{N}_{i=1}[Y_{i,1}(v_1)- \bar Y_{1}(v_1) ]^2, \\
S^2_{\Delta} & = \frac{1}{N-1}\sum\limits^{N}_{i=1}[(Y_{i,2}(c,c) - Y_{i,1}(c)) - (\bar Y_{2}(c,c) - \bar Y_{1}(c) ) ]^2, \\
S_{\tau}^2 & = \frac{1}{N-1}\sum\limits^{N}_{i=1}[Y_{i,2}(c,v_2) - Y_{i,1}(v_1) - (Y_{i,2}(c,c) - Y_{i,1}(c))- \tau_1]^2.
\end{aligned}
\end{equation*}
\end{thm} 
Notice that $S^2_{\tau}$ involves product terms of potential outcomes of the same units under different treatment paths and is generally not estimable. $S^2_{c,v_2}$, $S^2_{v_1}$ and $S^2_{\Delta}$ can be consistently estimated with their sample counterparts. For example, let
\begin{equation}
\begin{aligned}
s^2_{c,v_2} = \frac{1}{N_{c,v_2} - 1}\sum\limits_{i:\bZ_i = (c,v_2)} [Y_{i,2}^{\obs} - \widehat{\bar Y}_2(c,v_2)]^2, \quad  \widehat{\bar Y}_2(c,v_2) = \frac{1}{N_{c,v_2}}\sum\limits_{i:\bZ_i = (c,v_2)}Y_{i,2}^{\obs},
\end{aligned}
\label{eq:sample_var}
\end{equation}
and $\hat V_{\tau_1}= N_{c,v_2}^{-1}s^2_{c,v_2} + N_{v_1}^{-1}s^2_{v_1} + N_{c,c}^{-1}s^2_{\Delta}$.
\begin{thm}\label{thm:thm_2}
Under regularity conditions, the Wald-type interval $(\hat \tau_1 - z_{\alpha/2} \hat V_{\tau_1}^{1/2}, \hat \tau_1 + z_{\alpha/2} \hat V_{\tau_1}^{1/2})$ has asymptotic coverage rate $\geq 1-\alpha$, where $z_{\alpha/2}$ is the $(1-\alpha/2)$ quantile of a standard normal distribution.
\end{thm} 

It is important to note that the terms $\bar Y_{2}(c, c) $ and $\bar Y_{1}(c)$  in $\tau_1$ are estimated with the same samples which were in control in both iterations. Thus, although we could have a more accurate estimate of $\bar Y_{1}(c)$ by using all samples in the control group in the first iteration, this will result in a subset of samples also being used in the estimate of $\bar Y_{2}(c, v_2)$ and significantly complicates the variance estimation of $\hat\tau_1$.

When using an experiment to estimate the average treatment effect, practitioners usually focus on a single iteration (i.e., the max-power iteration) and only uses observations from that iteration in the estimate. On the contrary, $\hat\tau_1$, involves observations from multiple iterations. Although any direct comparison of potential outcomes in the same iteration has clear causal interpretations, this is not necessarily true for cross-iteration comparisons. As shown in the Online Supplementary Materials B, $\tau_1$ can be interpreted as the treatment effect of a special treatment under additional identifiability assumptions. Namely, we refer to the treatment underlying $\tau_1$ as the \textit{platform treatment} which captures the ability to implement iterative experiments vs. the control of no iterative experiments.


\subsection{Estimating VOIE with repeated max-power iterations}
\label{sec:single_2}

In Section~\ref{sec:single_1}, $\tau_1$ are estimated under progressive ramps with $p_1 < p_2$. In practice, it may be possible to repeat a certain iteration (usually the max-power iteration) and test the final version of the product to be launched, in which case $p_1 = p_2 = 0.5$. In this section, explain how to estimate $\tau_1$ under such designs.      

Since ramps before the max-power iteration are for mitigating risk, with little emphasis on treatment effect estimation, to avoid delaying the release, practitioners recommended that this stage is relatively short-lived if no severe degradations are detected \citep{xu2018sqr}. This is especially prominent when the experiment has deficient power. To overcome this, sometimes, practitioners recommend quickly ramping up a minimal viable product to the max-power iteration and using the feedback gathered at this stage to improve the feature. These improvements are required to be minor to reduce the likelihood of bringing additional risk. For example, typical improvements at the $50\%$ ramp include better code implementations that are tested offline, choosing different tuning parameters in machine learning models or even slight changes to the model itself given the feedback from previous iterations. After these non-risky changes are made, it is inefficient to go through the entire phased release cycle. Instead, an additional $50\%$ ramp is implemented to accurately estimate the treatment effect of the final version before launching to a larger group of users. 

\begin{figure*}
\vspace{0.1cm}
\begin{center}
\tikzstyle{level 1}=[level distance=3.5cm, sibling distance=3.3cm]
\tikzstyle{level 2}=[level distance=3.5cm, sibling distance=1.6cm]
 \begin{tikzpicture}[->, grow=right, sloped]
\node[root] {Start}
    child {
        node[bag] { $Y_{i,1}(c)$}        
            child {
                node[bag] {$Y_{i,2}(c,c)$}
                edge from parent[red, line width=0.5mm]
                node[above] {}
                node[below]  {$$}
            }
            edge from parent[red, line width=0.5mm]
            node[above] {$$}
            node[below]  {$$}
    }
    child {
        node[bag] { $Y_{i,1}(v_1)$}        
            child[dashed, gray] {
                node[unobs] {$Y_{i,2}(v_1,c)$}
                edge from parent
                node[above] {}
                node[below]  {}
            }
            child {
                node[bag] {$Y_{i,2}(v_1, v_2)$}
                edge from parent
                node[above] {$$}
                node[below]  {$$}
            }
            edge from parent 
            node[above] {$$}
            node[below]  {$$}    }
;
\end{tikzpicture}
\end{center}
\caption{Potential outcomes for unit $i$ under the two-iteration stepped-wedge design when  $p_1 = p_2$. The gray node represents the potential outcome when the feature is de-ramped after the first iteration. The red path illustrates one possible realization of the potential outcomes.
}
\label{fig:po2}
\end{figure*}
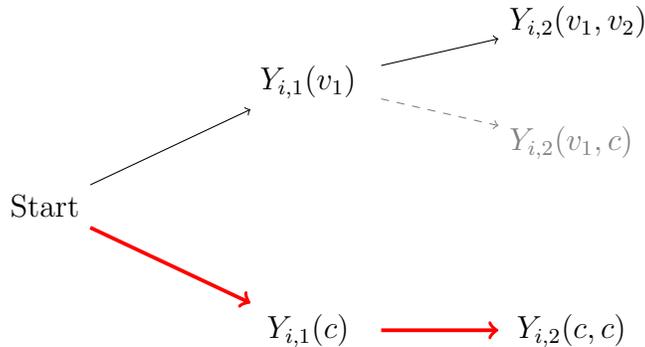

In this scenario, no changes are made until the first max-power iteration, which is picked as the LU. The second $50\%$ ramp, after incorporating the changes, is then the LMP. The potential outcomes under this design are illustrated in \ref{fig:po2}. From the product development perspective, the designs in Section~\ref{sec:single_1} and Section~\ref{sec:single_2} represent two different phased release strategies summarized in~\ref{tab:comparison}. The key difference between the two designs is that under the progressive iterations design, the risk mitigation stage and the product development stage are combined. 

To estimate $\tau_1$ under the repeated max-power iteration design, we require an additional assumption.
\begin{assumption}[No carryover effect]
\label{ass:carryover}
There is no carryover effect if $Y_{i,2}(c, v) = Y_{i,2}( v_1, v)$ for $v\in \{v_1,v_2\}$.
\end{assumption}
The validity of the no carryover effect assumption depends on the feature being tested and needs to be evaluated on a case-by-case basis. It is a good practice to allow a cool-down period, in which all units are assigned to control, between the two iterations to make this assumption more plausible. Under Assumption~\ref{ass:constant} and the no carryover effect assumption, $\tau_1$ in (\ref{eq:voie_ass}) becomes
\begin{equation*}
\begin{aligned}
\tau_1 & =  \frac{1}{N}\sum\limits^{N}_{i=1}\{[Y_{i, 2}(v_1, v_2) -  Y_{i, 2}(c,c)] -[Y_{i, 1}(v_1) -  Y_{i, 1}( c) ] \} .
\end{aligned}
\end{equation*}

\newcolumntype{M}[1]{>{\centering\arraybackslash}m{#1}}
\begin{table}[]
\centering
\begin{tabular}{p{7mm}p{41mm}M{50mm}M{50mm}}
\toprule
 & Stage                  & Progressive iterations                                 & Repeated MP iterations                \\ \midrule
$\RN{1}$  & Risk Mitigation     & \multirow{2}{*}{\vspace{-1.1cm} \shortstack{Iterations before the \\ $50\%$ ramp ($*$)} } & Iterations up to the first $50\%$ ramp       \\ \cmidrule(r){1-2} \cmidrule(l){4-4} 
$\RN{2}$  & Product Development &                                                        & The interim between the two $50\%$ ramps ($*$) \\ \midrule
$\RN{3}$  & Effect Estimation   & The $50\%$ ramp                                        & The second $50\%$ ramp                       \\ \bottomrule
\end{tabular}
\caption{Comparison of the phase release cycle under two designs. $*$ indicates that non-risky changes to the product are allowed at that stage.}
\label{tab:comparison}
\end{table}

Similar to (\ref{eq:plugin}), we consider the following plug-in estimator of $\tau_1$:
\begin{equation}
\begin{aligned}
\hat\tau_1 &= \frac{1}{N_{v_1,v_2}}\sum\limits^{N}_{i=1}[Y_{i,2}^{\obs} \mathbbm{1}(\bZ_i = (v_1, v_2)) - Y_{i,1}^{\obs}  \mathbbm{1}(Z_{i,1} = v_1)] \\
&\quad  - \frac{1}{N_{c,c}} \sum\limits^{N}_{i=1}[Y_{i,2}^{\obs} \mathbbm{1}(\bZ_i = (c, c)) - Y_{i,1}^{\obs} \mathbbm{1}(Z_{i,1} = c)],
\end{aligned}
\label{eq:plugin_2}
\end{equation}
where $N_{v_1,v_2} = \sum^{N}_{i=1} \mathbbm{1}(\bZ_i = (v_1, v_2))$ is the number of units in the treatment group, $N_{c,c}$ the number of units in the control group. Similar to Theorem~\ref{thm:thm_1}, we have 

\begin{thm}
\label{thm:thm_3}
Under the multiple max-power iteration stepped-wedge design $\eta$, the plug-in estimate in (\ref{eq:plugin_2}) satisfies 
\begin{equation*}
\begin{aligned}
\mathbbm{E}[\hat\tau_1] = \tau_1, \quad \mathbbm{V}[\hat\tau_1] = \frac{1}{N_{v_1,v_2}}S_{v_1,v_2}^2  + \frac{1}{N_{c,c}}S_{c,c}^2 - \frac{1}{N}S_{\tau}^2,
\end{aligned}
\end{equation*}
where 
\begin{equation*}
\begin{aligned}
S^2_{v_1,v_2} & = \frac{1}{N-1}\sum\limits^{N}_{i=1}[(Y_{i,2}(1;v_1,v_2) - Y_{i,1}(1;v_1)) - (\bar Y_{2}(v_1,v_2) - \bar Y_{1}(v_1)) ]^2, \\
S^2_{\Delta} & = \frac{1}{N-1}\sum\limits^{N}_{i=1}[(Y_{i,2}(1;c,c) - Y_{i,1}(1;c)) - (\bar Y_{2}(c,c) - \bar Y_{1}(c) ) ]^2, \\
S_{\tau}^2 & = \frac{1}{N-1}\sum\limits^{N}_{i=1}[Y_{i,2}(1;v_1,v_2) - Y_{i,1}(1;v_1) - (Y_{i,2}(1;c,c) - Y_{i,1}(1;c))- \tau_1]^2.
\end{aligned}
\end{equation*}
Let $\hat V_{\tau_1}= N_{v_1,v_2}^{-1}s^2_{v_1,v_2} + N_{c,c}^{-1}s^2_{\Delta}$, Theorem~\ref{thm:thm_2} also holds.
\end{thm}


\section{Quantifiable value of an experimentation platform}
\label{sec:platform}

Typically, the primary goal metric $Y$ can be hard to move, especially when $Y$ is a company-level metric such as the overall revenue or the total page views of the website. Therefore, the VOIE $\tau_1$ is likely to be small in absolute value, and so inference from a single experiment would have low power, especially under the design in Section~\ref{sec:single_1} where treatment allocation in the first iteration is small. Fortunately, from an operational perspective, firms are more interested in the aggregated value of multiple experiments at the team or platform level. 

To this end, let $E_j$, $j=1,2,\ldots, J$ denote a set of iterative experiments whose values we want to aggregate. We assume that the experiments all have the same target population and are mutually independent. Let $\tau_{1,j}$ be the value from the $j$-th iterative experiment. We define the aggregated value of these experimentations as
\begin{equation}
\begin{aligned}
\delta_{\bm{a}} = \sum\limits^{J}_{j=1}a_j\tau_{1,j},
\end{aligned}
\end{equation}
where $\bm{a} = (a_1,a_2,\ldots,a_J)$, $\sum^{J}_{j=1}a_j=1.$ For fixed $\bm{a}$, the plug-in estimator $\hat\delta_{\bm{a}}$ is unbiased and has the following estimated upper bound of its variance $\hat{\mathbbm{V}}[\hat\delta_{\bm{a}}]=\sum^{J}_{j=1}a_j^2\hat{\mathbbm{V}}[\hat\tau_{1,j}]$. The weights $\bm{a}$ can be selected based on the \textit{a priori} evaluated importance of the tested features. When no such prior knowledge is available, a reasonable default is to estimate the inverse-variance weighted average 
\begin{equation}
\begin{aligned}
\delta_{\text{inv}} = \sum\limits^{J}_{j=1}(\sum\limits^{J}_{j=1}\mathbbm{V}[\tau_{1,j}]^{-1})^{-1}\mathbbm{V}[\tau_{1,j}]^{-1}\tau_{1,j},
\end{aligned}
\label{eq:voe_agg}
\end{equation}
which has the smallest variance among all $\delta_{\bm{a}}$'s. To estimate $\delta_{\text{inv}}$, we have to treat the estimated upper bounds of $\mathbbm{V}[\hat\tau_{1,j}]$ as its true variance. When the bounds are tight for each experiment (or at least tight for those experiments with large weights), this gives a reasonable approximation to $\delta_{\text{inv}}$ due to the large sample size in online experimentations. Note that the definition of $\delta_{\bm{a}}$ requires all experiments to have the same target population such that their values are comparable. If this is not the case, the VOIE can be computed and aggregated at the site-wide impact level \citep{xu2015infrastructure}.

Although the definition of $\delta_{\bm{a}}$ is simple, it gives rise to a set of metrics by varying the primary goal metric, $Y$, and the set of experiments under consideration. These metrics provide a holistic picture of the learning behavior across teams and help identify weaknesses and opportunities. In practice, $\delta_{\bm{a}}$ can be defined at different granularity levels and estimated on a monthly or quarterly basis. For example, computing  $\delta_{\bm{a}}$ at the platform level lets us monitor the contribution of iterative experimentations to core metrics. The aggregated VOIE on these metrics also serve as direct performance metrics of the experimentation platform itself. As another example, for each core metric, we can compute the aggregated VOIE at the organization or team level and identify which teams benefit the most from iterative experimentation. 


\section{Aggregated value of LinkedIn's experiment platform}
\label{sec:application}

LinkedIn uses an internally developed, unified, and self-served \textit{Targeting, Ramping and Experimentation Platform} (T-REX) for iterative experimentations on a user population of over 700 million members worldwide. The platform has been evolving in the past decade from a small experiment management system with limited functions into a full-fledged platform that allows efficient iterative experimentation and can serve more than 40000 tests simultaneously at any given time \citep{xu2015infrastructure, trexBlog}. During this evolution, numerous features and improvements have been incorporated into the platform include infrastructure changes that accelerate the experiment velocity, optimizations of the offline analysis data pipeline, supports of more advanced statistical methodologies for experimentation readout, and user interface changes that improve developers' experience with the platform. 

LinkedIn has a culture to ``test everything" --- essentially, all new features have to go through a ``ramp up" process enabled by the T-REX platform, which provides instant feedback to help make product launch decisions. It is well acknowledged that T-REX provides value in this regard, but there is little understanding of what that value is and how it changes over time. On the other hand, although T-REX serves all other teams, it lacks a reliable way to collect feedback on the impact of feature changes on the platform itself. Traditionally, the team has been relying on indirect metrics such as the average experimentation duration or user surveys to quantify the performance of the T-REX platform, which are inefficient and non-standardized. 
\begin{figure}[ht!]
\begin{center}
\centerline{\includegraphics[width=0.9\columnwidth]{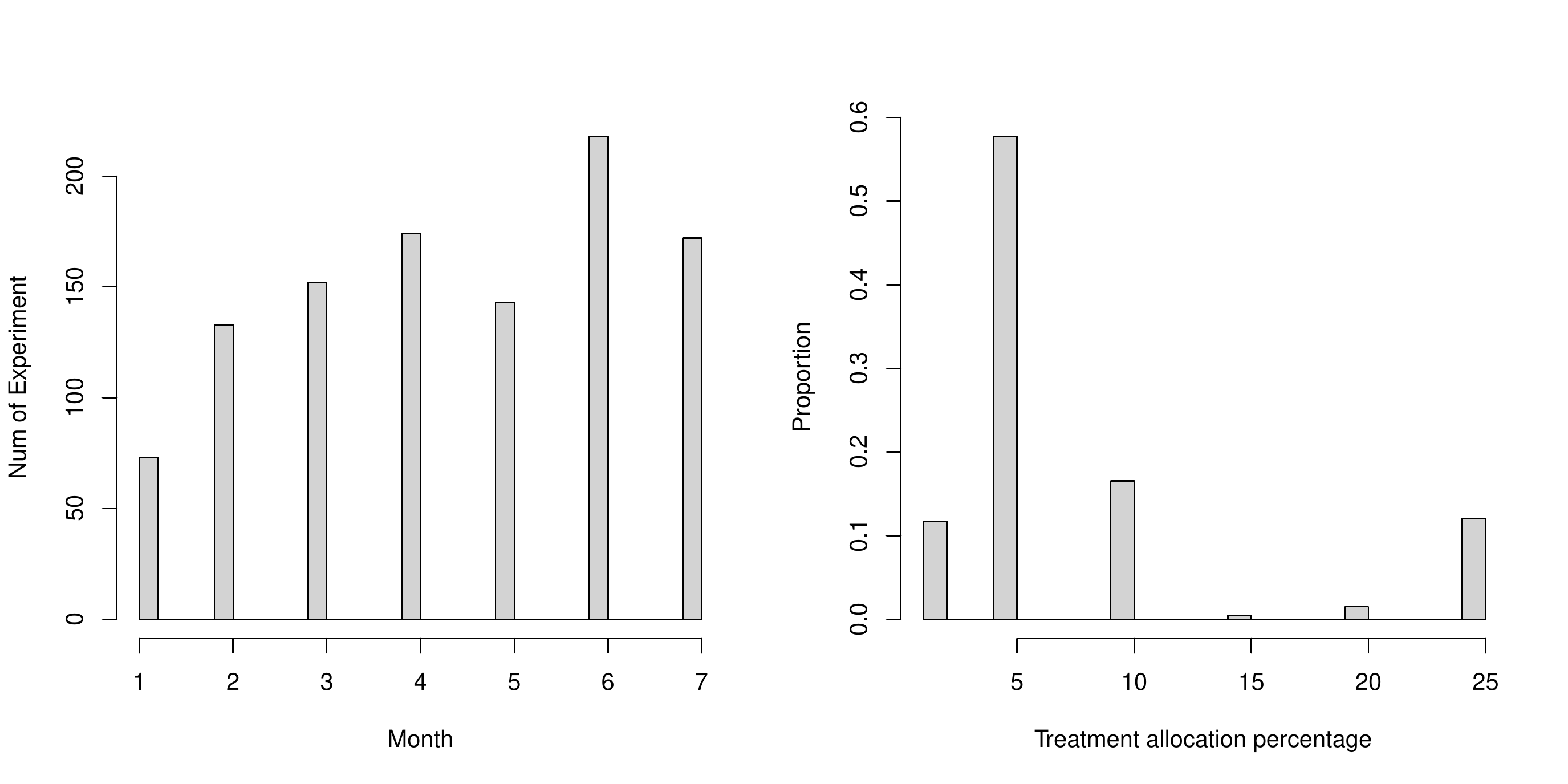}}
\caption{Left: number of experiments by end month; Right: number of experiments by the treatment allocation percentage in LU.}
\label{fig:hist_size}
\end{center}
\end{figure}

In this work, we recommend monitoring the company (platform) level VOIE as a natural performance metric that captures the efficiency of the platform and its ability to guide product development. This estimand also helps to understand the pattern of learning at the company-level. As an illustration, we estimate the aggregated monthly VOIE across all experiments completed between January 2019 and July 2019 at LinkedIn for a primary metric measuring overall user engagement. Specifically, we consider all experiments in which the treatment was ramped to $50\%$ during the analysis period. For each experiment, we pick its first iteration as the LU and its last $50\%$ iteration as the LMP to ensure that no product changes can happen before the LU or after the LMP. The VOIE of each experiment is computed and then aggregated based on (\ref{eq:voe_agg}) to get the company-level VOIE. In our example, we only considered experiments with at least 10000 samples and lasted for at least three days in both iterations to filter out experiments with a very small target population or stringent triggering conditions. This step is optional as the experiments that were filtered out would likely have small weights in (\ref{eq:voe_agg}). There are $1065$ experiments under consideration after filtering. 

\ref{fig:hist_size} (left) shows the number of experiments that finished in each month, which increases steadily in the first quarter of 2019 and fluctuates afterward, yielding an average of 152 experiments per month. \ref{fig:hist_size} (right) summarizes the treatment percentage in the LU iteration of the experiments. Most experiments follow the company's recommendation to start with small ramps with ramping percentage in $\mathcal{P} = \{1\%, 5\%, 10\%$, $25\%\}$. For experiment iterations that last for more than two weeks, we only use their results in the first two weeks to focus on the short-term treatment effect and limit the scope of uncontrolled time effects to make the VOIE across experiments more comparable.

\begin{figure}[ht!]
\begin{center}
\includegraphics[width=0.45\columnwidth]{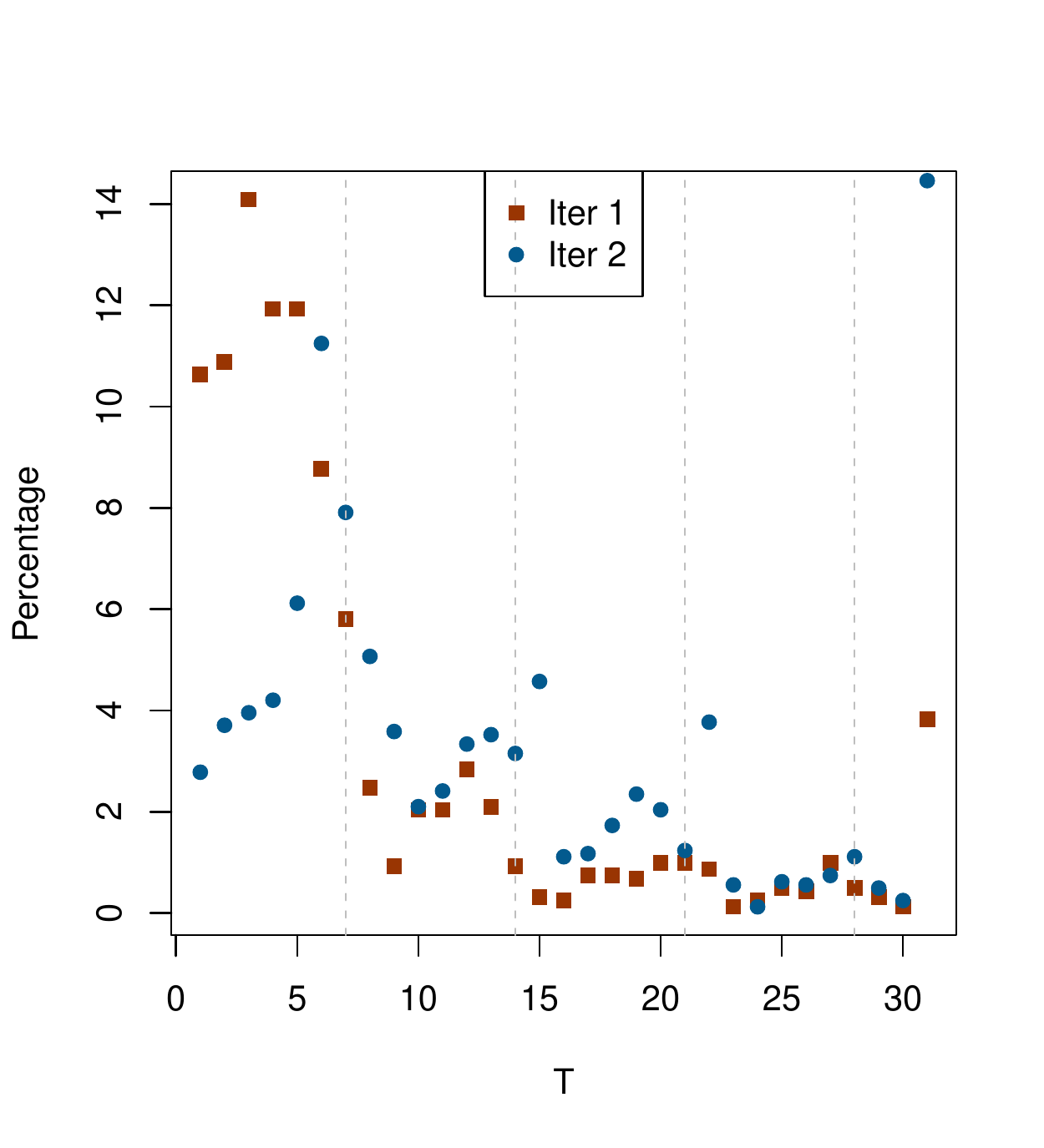}\includegraphics[width=0.45\columnwidth]{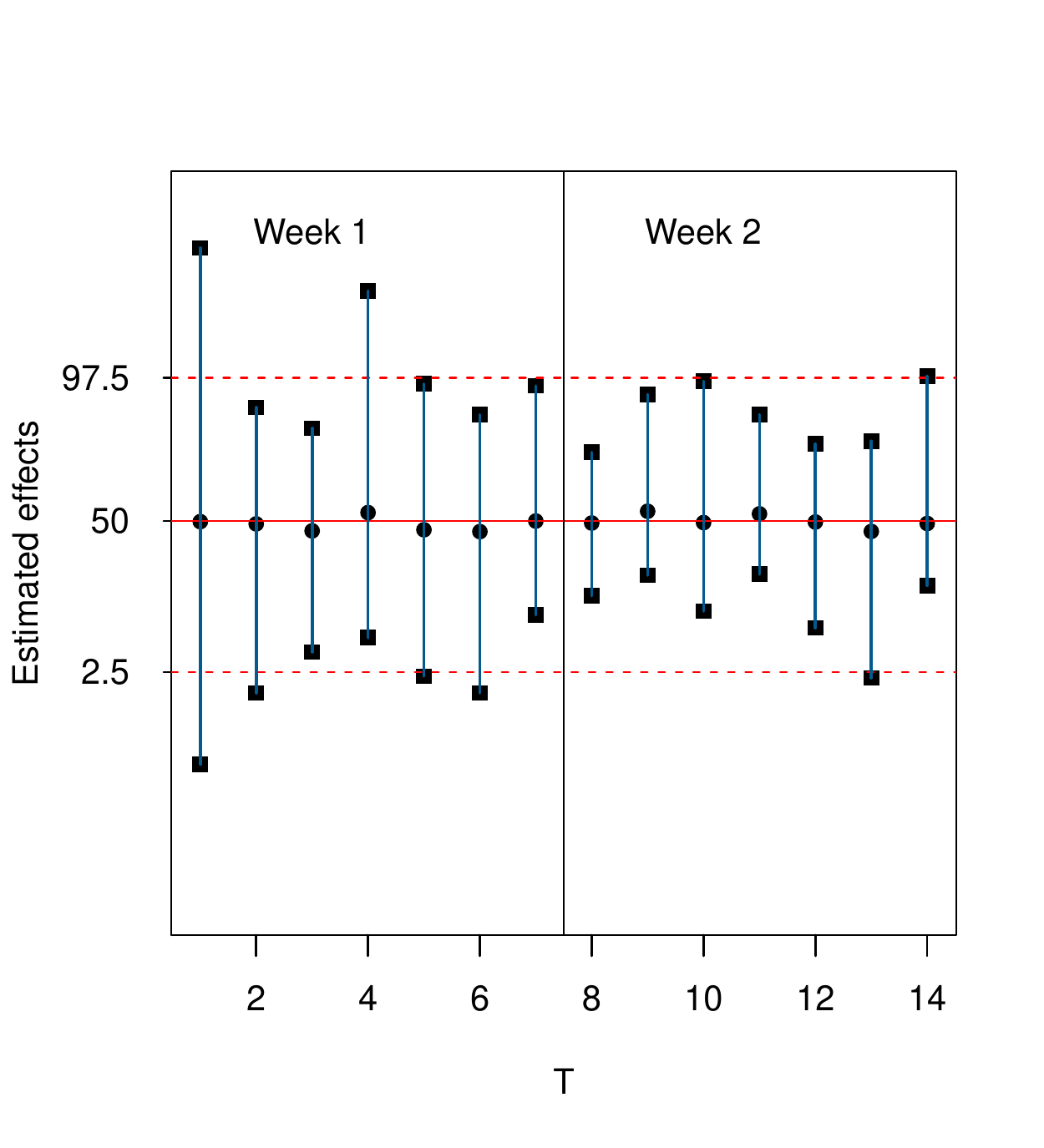}
\caption{Left: Relation of the number of experiments and the experiment duration $T$ (the number of experiments are shown as the percentage of the total number of experiments); Right: Relation between the estimated daily effects and the day in the experiment. The dashed lines mark the $2.5$ and $97.5$ quantiles of the estimated effects across all experiments.}
\label{fig:count_duration}
\end{center}
\end{figure}

Let $\hat\delta_{\text{inv}}$ be the overall estimated VOIE and let $y_{2018}, y_{2019}$ be the per member daily average of the core metric in the first 7 months of 2018 and 2019, respectively. Let $f(x) = x/(y_{2019}- y_{2018})$, we have $f(\hat\delta_{\text{inv}}) = 0.203$. Since $y_{2019}- y_{2018} > 0$, the aggregated VOIE amounts to $20.3\%$ of the YoY increase in this metric in the first half of 2019. Note that $f(\hat\delta_{\text{inv}})$ is not of special interest \textit{per se}, we use it to encrypt the raw metric values per company's policy. Moreover, the \textit{p}-value for testing $H_0: \delta_{\text{inv}}=0$ vs. $H_1: \delta_{\text{inv}}\not=0$ with $t$-test and an estimate of ${\mathbbm{V}}[\hat\delta_{\text{inv}}]$ based on the estimated upper bounds of ${\mathbbm{V}}[\hat\tau_{1,j}]$ is 0.068. 

It is important to note the underlying time component in the definition of VOIE in (\ref{eq:voie}) inherited from the definition of the potential outcomes. Specifically, $\tau_{1,j}$ is defined as the daily average value of experiment $j$. Since different experiments have different running times, their values are estimated based on different time ranges. The duration of the experiments are shown in \ref{fig:count_duration} (left). Typically, the LU iteration of most experiments finishes within a week and has a much shorter running time than the LMP iteration, as expected. \ref{fig:count_duration} (right) shows the quantiles of the estimated treatment effects at the $T$-th day of the experiment for $1\leq T\leq 14$. For example, the bar at $T = 8$ shows the observed $(2.5\%, 97.5\%)$ quantiles of the estimated effects on the 8-th day of the iteration. Based on this plot, the estimated treatment effects do not have a clear trend over time at the platform level. Thus there is no sign of significant heterogeneity of VOIE by experiment duration, and the overall VOIE $\hat\delta_{\text{inv}}$ serves as a good summary of the value of experimentation at the platform level.

\begin{figure}[ht!]
\begin{center}
\includegraphics[width=0.33\columnwidth]{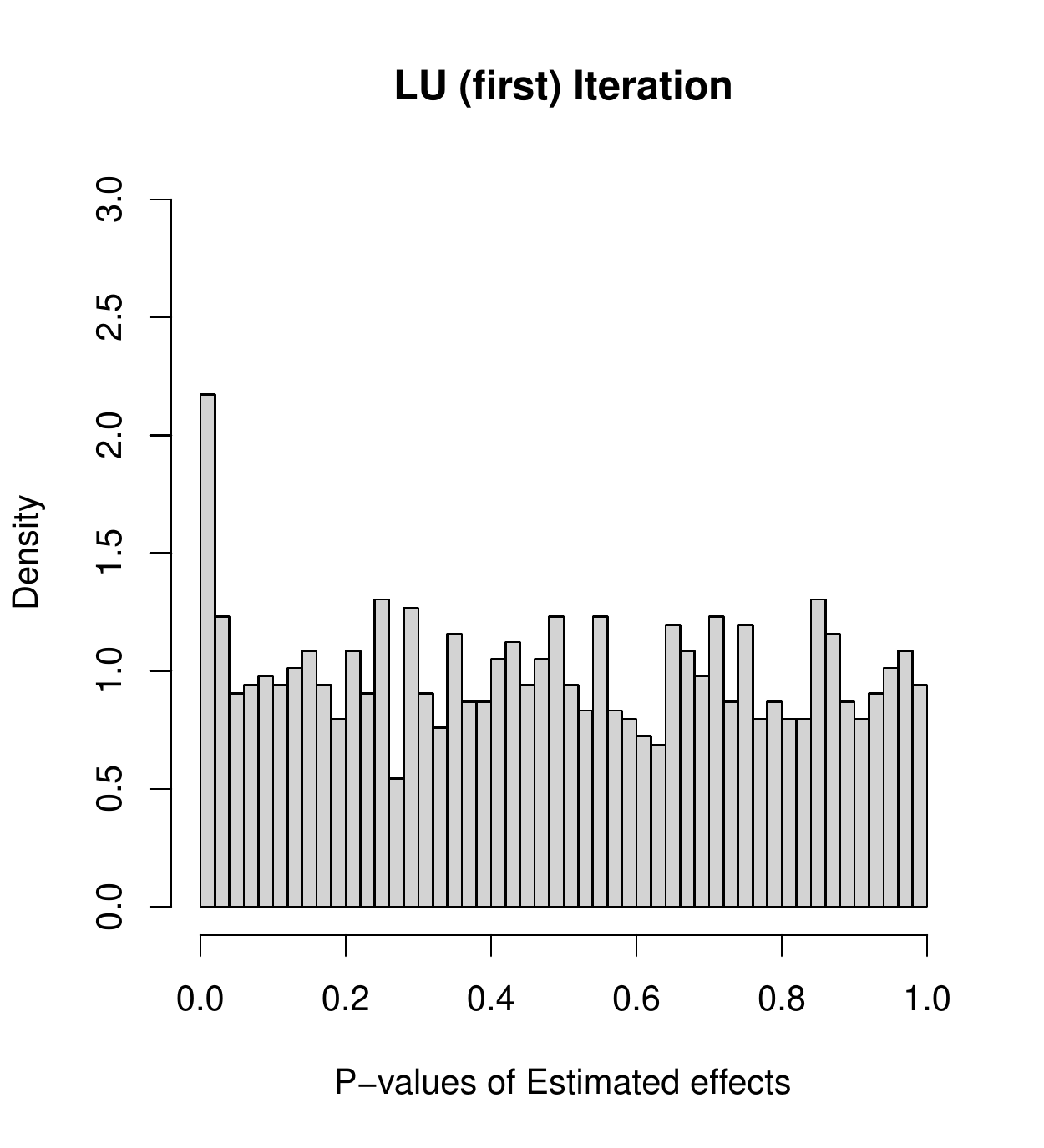}\includegraphics[width=0.33\columnwidth]{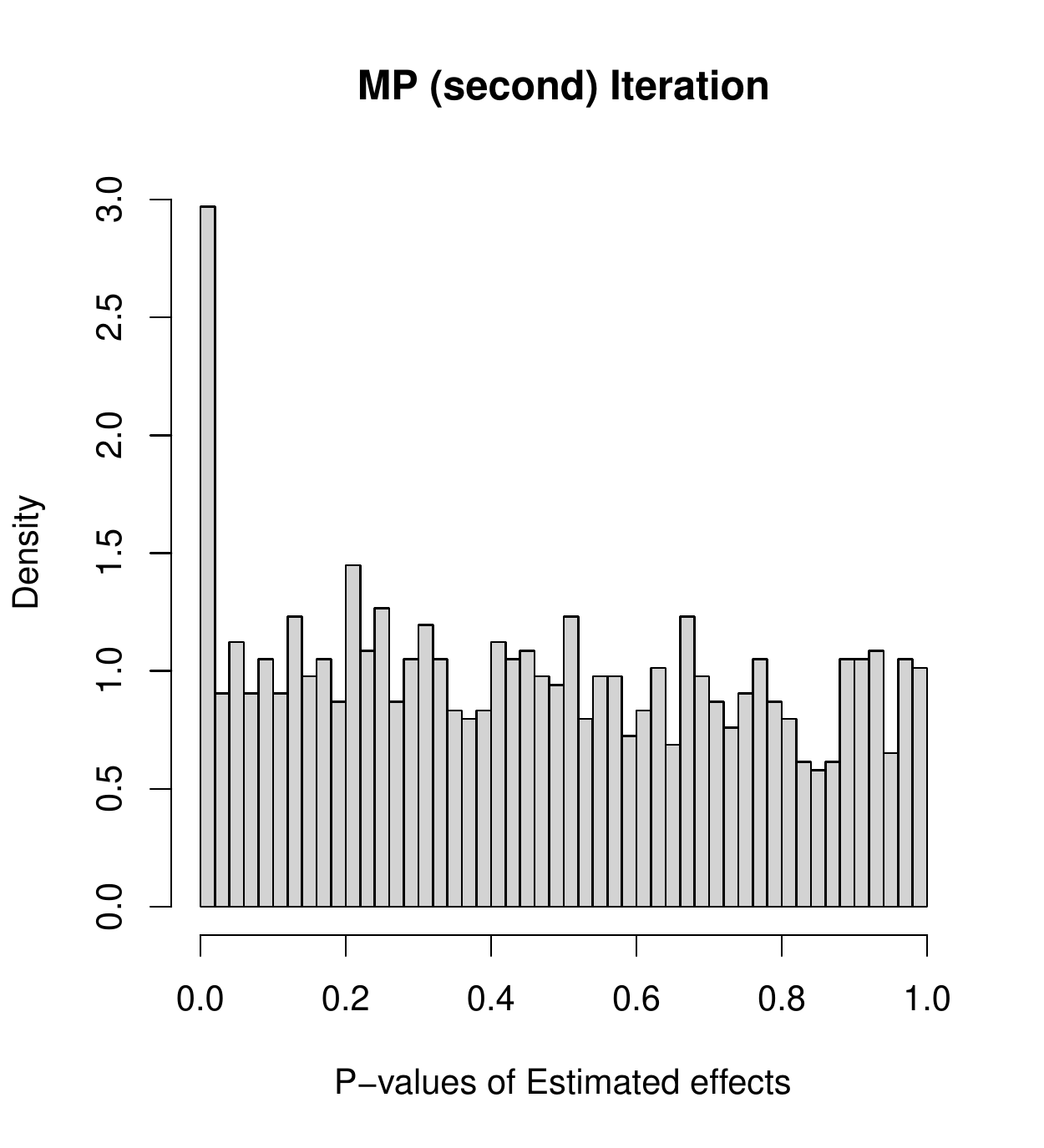}\includegraphics[width=0.37\columnwidth]{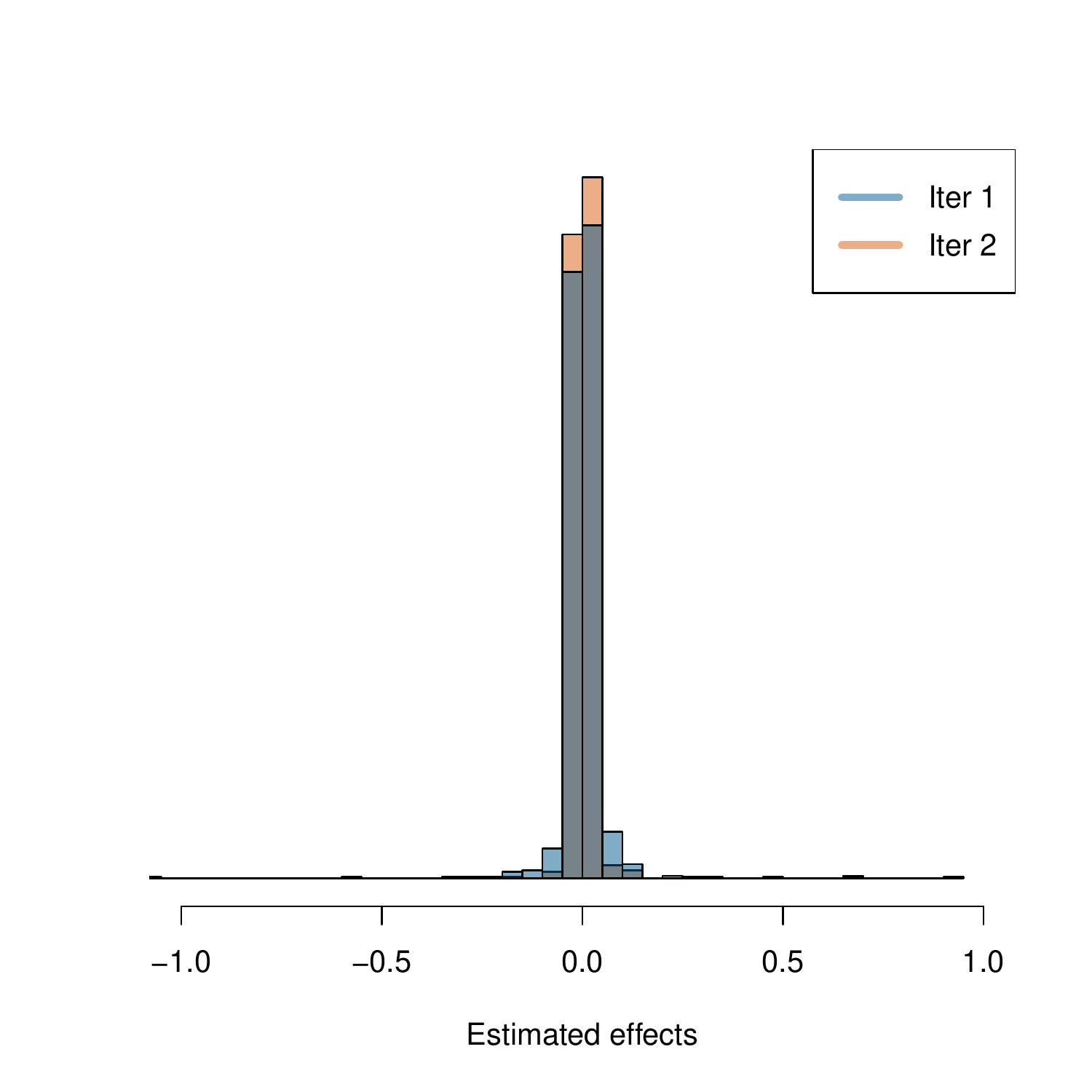}
\caption{Left and middle: $p$-values of estimated treatment effects in the LU and the LMP iteration; Right: estimated effects from all experiments in the two iterations.}
\label{fig:hist_pval}
\end{center}
\end{figure}

\ref{fig:hist_pval} shows the \textit{p}-values and the estimated treatment effects in the two iterations of all experiments in the analysis. As expected, the LMP iteration tends to show more statistically significant results than the LU iteration and there are more positive effects than negative ones in both iterations. In either iteration, the \textit{p}-value distribution is a mixture of a spike close to 0 and an almost uniform slab. This mixture reflects different roles of the metrics in the experiments. For a small number of the features, this engagement metric is the primary goal metric they hope to move. Experiments that test these features are more likely to see smaller \textit{p}-values. However, in most experiments, this metric serves as a guardrail metric on which the treatment effect is expected to be neutral. The aggregated VOIE $\hat\delta_{\text{inv}}$ captures the VOIE on all experiments regardless of the role of the metric and is thus diluted. A triggered analysis can be performed to remove this dilution effect by only looking at experiments that treat the metric as the primary goal metric.

\begin{figure}[ht]
\begin{center}
\includegraphics[width=0.45\columnwidth]{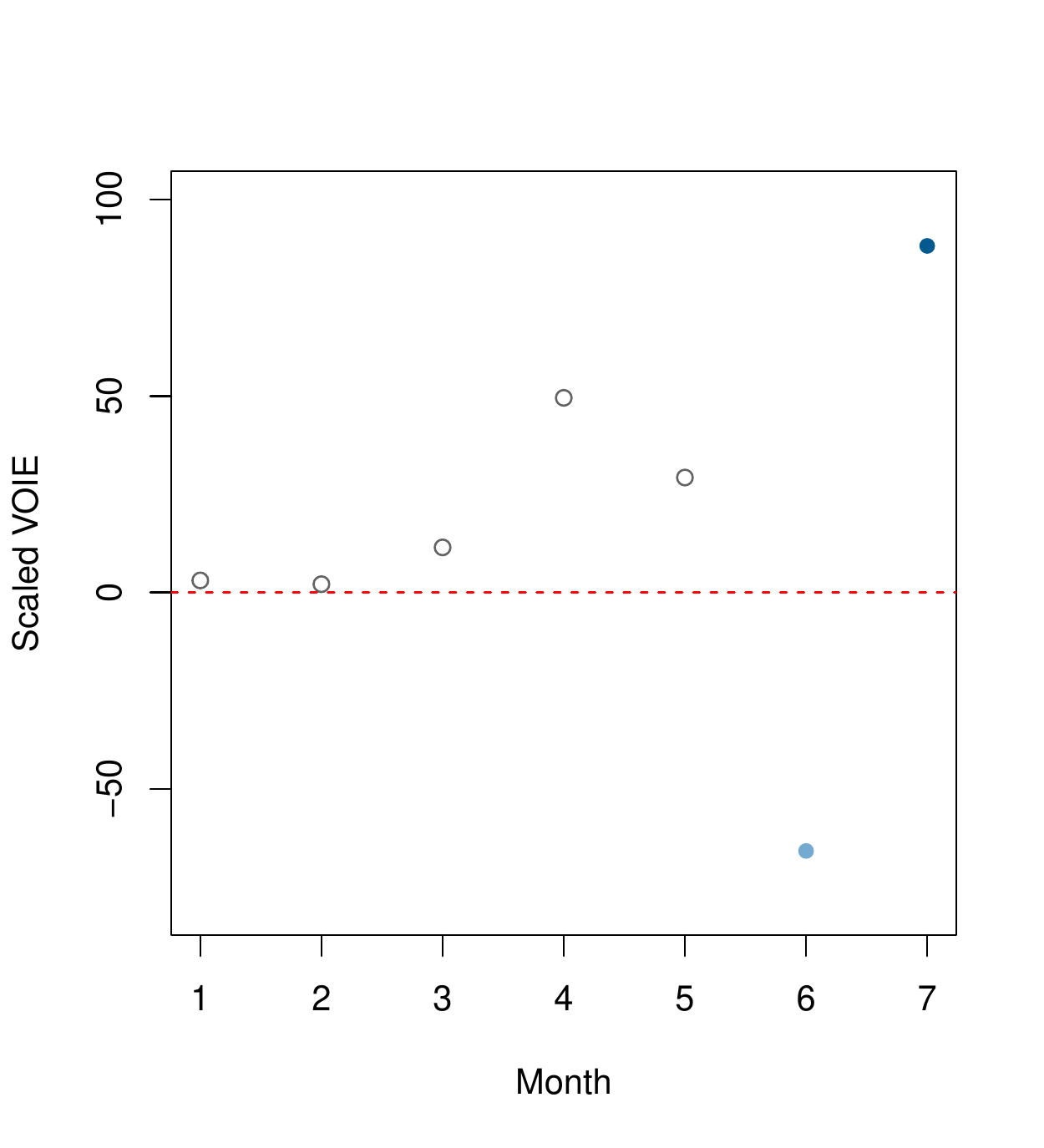}\includegraphics[width=0.45\columnwidth]{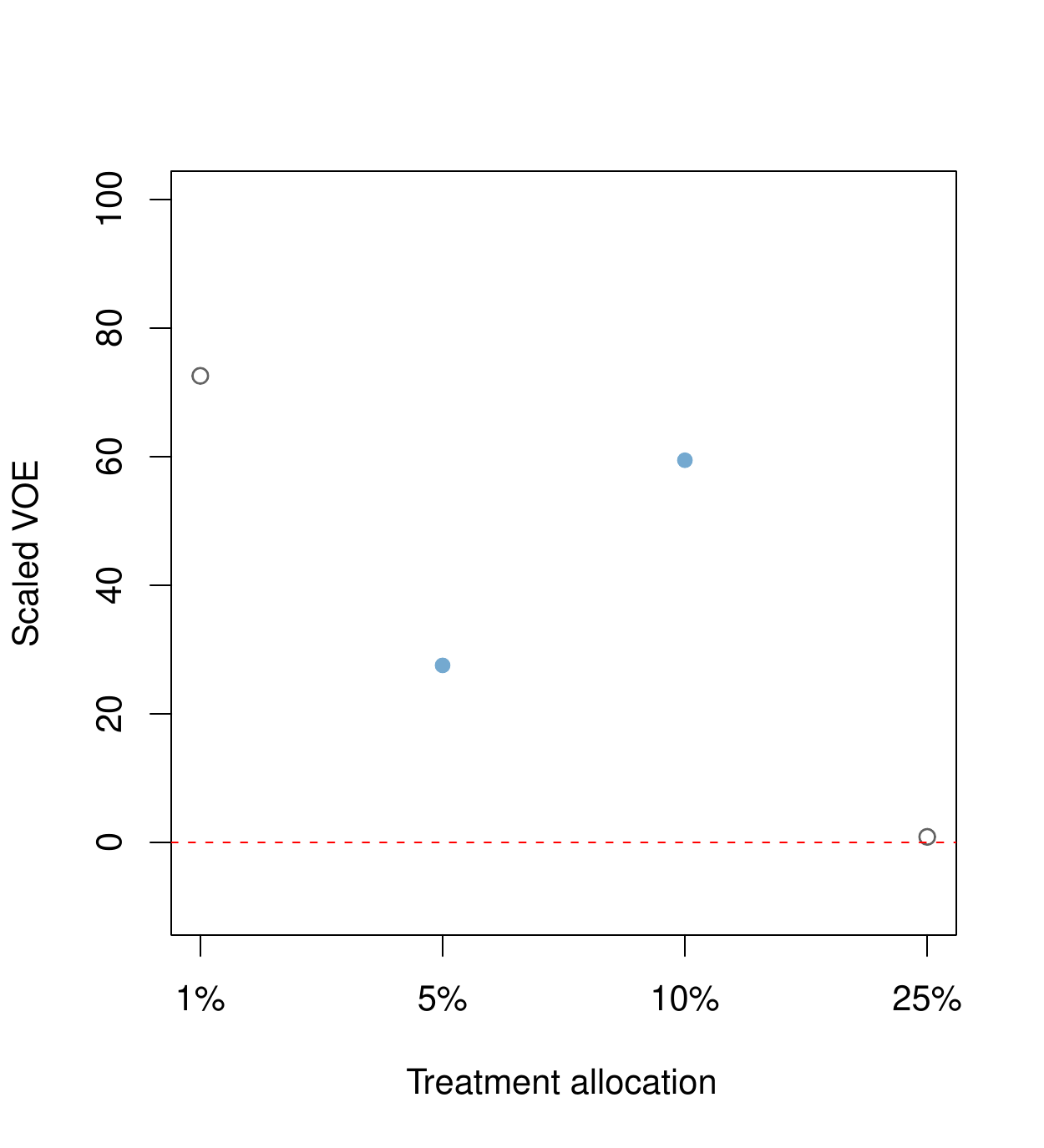}
\caption{Left: VOIE by end month; Right: VOIE by treatment allocation in LU. All values are transformed from the raw VOIE by $f(\cdot)$. Light blue marks estimates that are significant at the $0.05$ level; dark blue marks estimates that are significant at the $0.01$ level.}
\label{fig:voe}
\end{center}
\end{figure}
 
To see the trend of company level learning, we compute the aggregated VOIE of experiments finished in each month and plot the results in \ref{fig:voe} (left). In the first half of 2019, the VOIE shows an increasing trend and is positive in all months but June. There are likely two reasons for the unusual behavior of VOIE in June and July. First, there is a company shutdown in the 4th of July week and the moratorium policy that no feature modifications are allowed one week previous to the shutdown. As a result, feature improvements that would have been made in June have to be pushed back in July, leading to an extremely high VOIE in that month. Second, June and July is the transitional period of LinkedIn to a new fiscal year. During this period, many metrics, including the one we analyzed here, are not stable due to various reasons related to budgeting and customer spending.

Similarly, we also aggregate the VOIE of experiments according to their treatment allocation in the LU iteration. The results are shown in \ref{fig:voe} (right). Since we take the first iteration of an experiment as its LU, the treatment allocation in this iteration reflects the experimenter's initial evaluation of the risk of the feature, which also serves as a proxy for the feature's room for improvement. For example, features that start the ramp at $1\%$ tend to have larger risks, either because the features are premature or the engineering implementations are unpolished. During the iterative process, both the features and their implementations can be improved. These features are more likely to benefit from iterative experimentation and thus have a larger VOIE. On the other hand, features that were directly ramped up to $25\%$ are believed to be safe and closer to completion. For these features, the iterative process is usually shorter, and their VOIE tends to be close to zero.


 \section{Concluding remarks}
 \label{sec:conclude}

We have introduced a novel way to quantify the value of an iterative experiment and the overall value of an experimentation platform. VOIE has several attractive properties: (1) it directly quantifies the value in terms of certain business metrics. Thus no transformation of the metrics is required to interpret the results; (2) it only relies on the data that has already been collected and does not require collecting any new information. Therefore, the cost of implementing VOIE on top of a mature experimentation platform is negligible; (3) although not discussed in this paper, our strategy of defining VOIE can be extended to more complex designs with multiple iterations. How to estimate VOIE under various interference structures such as in the social network setting or the bipartite marketplace setting \citep{pouget2019testing, doudchenko2020causal, liu2021trustworthy} is an interesting direction to explore. Moreover, since we formulate iterative experiment in a general way that directly accounts for its multiple iteration nature, other causal estimands that depend on potential outcomes in multiple iterations can be defined and estimated under the same framework.


\section*{Acknowledgement}
We would like to thank Shan Ba, Min Liu and Kenneth Tay for their careful review and feedback; We would also like to thank Parvez Ahammad and Weitao Duan for their continued support.

\bibliography{voe-bib}

 
\clearpage
 \appendix
 
\pagenumbering{arabic}
\renewcommand*{\thepage}{S\arabic{page}}
 
 \beginsupplement
 
\section*{Supplementary Materials}

\renewcommand{\thesubsection}{\Alph{subsection}}

\setcounter{assumption}{0}
\renewcommand{\theassumption}{S\arabic{assumption}}

\renewcommand{\thefigure}{S\arabic{figure}}

\subsection{Proofs}


We provide proof of Theorem 1 and 2. Theorem 3 can be proved in a similar manner. At the second iteration, units in the population can be divided into three non-overlapping buckets based on their treatment paths. Specifically, let $\xi_i\in \{1,2,3\}$ be the bucket indicator of unit $i$ and define $\xi_i$ as  
\begin{equation*}
\begin{aligned}
\xi_i = \begin{cases}
1, \quad \text{ if } Z_{i,1} = v_1, \\
2, \quad \text{ if } \bZ_i = (c, v_2), \\
3, \quad \text{ if } \bZ_i = (c, c).
\end{cases}
\end{aligned}
\end{equation*}
Let $\eta$ denote the two-iteration stepped-wedge design. Instead of viewing $\eta$ in a sequential manner, we can view it as a ``collapsed" design with three treatment variants. In this design, $\xi_i$ is the treatment label of unit $i$. We also define the potential outcomes associated with this ``collapsed" design as $\mathcal{Y}_i(\xi_i = 1) = Y_{i, 1}( v_1)$, $\mathcal{Y}_i(\xi_i = 2) = Y_{i, 2}(c, v_2)$ and $\mathcal{Y}_i(\xi_i = 3) = Y_{i, 2}(c, c) - Y_{i, 1}(1, c) = \Delta_i$. We refer to this imaginary design as the auxiliary design and denote it as $\tilde\eta$. In our case, $\eta$ is implemented via two completely randomized designs. As a result, $\tilde\eta$ is a completely randomized design that randomly assigns $N_1 = p_1N$, $N_2=(1-p_1)p_2 N$ and $N_3 = (1-p_1)(1-p_2) N$ units to the three treatment buckets, respectively. Under $\tilde\eta$, we can write $\tau_1$ as
\begin{equation}
\begin{aligned}
\tau_1 = \frac{1}{N}\sum\limits^{N}_{i=1}[\mathcal{Y}_i(\xi_i = 2) - \mathcal{Y}_i(\xi_i = 1) - \mathcal{Y}_i(\xi_i = 3)].
\end{aligned}
\label{eq:trans}
\end{equation}
It is natural to consider the following plug-in estimator of $\tau_1$:
\begin{equation}
\begin{aligned}
\hat\tau_1 &= \sum\limits^{N}_{i=1}\sum\limits^{3}_{k=1}(-1)^{|2-k|}\frac{1}{N_k}\mathbbm{1}(\xi_i = k) \mathcal{Y}_i^{\rm{obs}},
\end{aligned}
\label{eq:plugin2}
\end{equation}
where $\mathcal{Y}_i^{\rm{obs}}$ is the observed outcome of unit $i$ under $\tilde\eta$. Inference is straightforward once noticing that samples used to estimate the three terms in (\ref{eq:trans}) do not overlap. In particular, we have
\begin{lem}
Let $ \bar{\mathcal{Y}}( k) = \frac{1}{N}\sum\limits^{N}_{i=1}\mathcal{Y}_i(\xi_i = k)$ for $k=1,2,3$. Under the two-iteration stepped-wedge design $\eta$, the plug-in estimate in (\ref{eq:plugin2}) satisfies 
\begin{equation*}
\begin{aligned}
\mathbbm{E}[\hat\tau_1] = \tau_1, \quad \mathbbm{V}[\hat\tau_1] = \sum\limits^{3}_{k=1} \frac{1}{N_k}S_k^2 - \frac{1}{N}S_{\tau}^2,
\end{aligned}
\end{equation*}
where 
\begin{equation*}
\begin{aligned}
S^2_k & = \frac{1}{N-1}\sum\limits^{N}_{i=1}[\mathcal{Y}_i(\xi_i = k) -  \bar{\mathcal{Y}}(k) ]^2, \quad k = 1,2,3, \\
S_{\tau}^2 & = \frac{1}{N-1}\sum\limits^{N}_{i=1}[\mathcal{Y}_i(\xi_i = 2) - \mathcal{Y}_i(\xi_i = 1) - \mathcal{Y}_i(\xi_i = 3) - \tau_1]^2.
\end{aligned}
\end{equation*}
\end{lem}
The proof of this lemma is straightforward following Theorem 3 in \cite{li2017general} under regularity conditions. Theorem 2 also follows from Theorem 5 in the same reference.

\subsection{Causal interpretation of VOIE}

Formally, let $W\in\{0,1\}$ where $W=1$ and $W=0$ indicate the counterfactual status with and without the iterative experiment platform, respectively. We refer to $W$ as the \textit{platform treatment} to differentiate it from the \textit{product treatment} $Z$. Let $W_{i,t} \in \{0,1\}$ be the platform treatment on unit $i$ at iteration $t$. Unlike the product treatment which can change over time, we view the platform treatment as fixed. This requires that no changes can be made to the experimentation platform during the experiment. Moreover, since the company can either have the platform or not, every unit must receive the same platform treatment. Formally, we have
\begin{assumption}[Constant platform treatment]
\label{ass:fixed}
    For $t,t^\prime=1,2$ and $i, i^\prime\in [N]$, $W_{i,t} = W_{i^\prime,t^\prime}$. We write the constant platform treatment as $W \in \{0, 1\}$.
\end{assumption}
With this additional treatment, we rewrite the potential outcomes of unit $i$ as $Y_i(W;\bZ_{1:N})$ and slightly modify the identifiability assumptions as follows:
\begin{assumption}[Non-anticipation]
\label{ass:nonanticipation2}
    The potential outcomes are non-anticipating if for all $i\in [N]$, $Y_{i,1}(w; \bz_{1:N, 1}, \bz_{1:N, 2} ) = Y_{i,1}(w; \bz_{1:N, 1}, \tilde\bz_{1:N,2})$ for all $w\in\{0,1\}$, $\bz_{1:N, 1} \in \{c, v_1 \}^N$, and $\bz_{1:N, 2},\tilde\bz_{1:N, 2} \in \{c, v_2\}^N$.
\end{assumption}
\begin{assumption}[No-interference]
\label{ass:nointerference2}
    The potential outcomes satisfy no-interference if for all $i\in[N]$, $t=1,2$ and $w=0,1$, $Y_{i,t}(w; \bz_{1:(i-1)}, \bz_i, \bz_{(i+1):N}) = Y_{i,t}(w; \tilde\bz_{1:(i-1)}, \bz_i, \tilde\bz_{(i+1):N})$ for all $\bz_{1:(i-1)},\tilde\bz_{1:(i-1)}\in\mathcal{V}^{i-1}$ and $\bz_{(i+1):N},\tilde\bz_{(i+1):N}\in\mathcal{V}^{N-i}$.
\end{assumption}
With these assumptions, the potential outcomes of unit $i$ become $Y_{i, 1}(W; Z_{i, 1})$ and $Y_{i, 2}(W; Z_{i, 1}, Z_{i, 2})$. We can now define VOIE as the treatment effect of the platform treatment:
\begin{definition}[Value of iterative experiment]
\label{def:voe2}
Let $\delta_{i, 1}(W = 0) = Y_{i, 1}(0; v_1) -  Y_{i, 1}(0; c)$ and $\delta_{i, 1}(W = 1) = Y_{i, 1}(1; v_2) -  Y_{i, 1}(1; c)$. We define the \textit{value of iterative experiment (VOIE)} on unit $i$ as $\tau_{i, 1} = \delta_{i, 1}(W = 1) - \delta_{i, 1}(W = 0)$. We also define the population average VOIE as 
\begin{equation}
\begin{aligned}
\tau_{1} =\frac{1}{N}\sum\limits^{N}_{i=1}\tau_{i,1}= \frac{1}{N}\sum\limits^{N}_{i=1}[\delta_{i, 1}(W = 1) - \delta_{i, 1}(W = 0)].
\end{aligned}
\label{eq:voe}
\end{equation}
\end{definition}
To estimate the population level VOIE  $\tau_1$, we make the following assumption:
\begin{assumption}[Full unawareness]
\label{ass:unawareness}
   The units are fully unaware of the experimentation if $Y_{i,t}\independent W$ for $i\in[N]$ and $t=1,2$.
\end{assumption}
 
\begin{assumption}[Time-invariant treatment effect]
\label{ass:constant_app}
   The individual-level treatment effects are time-invariant if $Y_{i,2}(W; c, v) - Y_{i,2}(W; c, c) = Y_{i, 1}(W; v) - Y_{i,1}(W; c)$ for $i\in [N], v\in \{v_1,v_2\}$ and $W = 0, 1$.
 \end{assumption}
 
Compared with the results in Section 3, we need an additional full unawareness assumption to give VOIE a formal causal interpretation.The full unawareness assumption never holds exactly as the experimentation platform adds latency to service loading. However, such delay is typically controlled to a level such that the users' experience is not harmed and can be ignored in practice. In online controlled experiments, the experimental units are always blind of the treatment assignment and usually have no information to tell whether they are in an experiment or not. Therefore, the full unawareness assumption is mild in our context. Under these assumptions, we can write $\tau_1$ as
\begin{equation}
\begin{aligned}
\tau_1 &= \frac{1}{N}\sum\limits^{N}_{i=1}\{[Y_{i, 1}(1; v_2) -  Y_{i, 1}(1; c)] -[Y_{i, 1}(0; v_1) -  Y_{i, 1}(0; c) ] \} \\
& =  \frac{1}{N}\sum\limits^{N}_{i=1}\{[Y_{i, 2}(1; c, v_2) -  Y_{i, 2}(1; c,c)] -[Y_{i, 1}(1; v_1) -  Y_{i, 1}(1; c) ] \} .
\end{aligned}
\label{eq:voe_id}
\end{equation}
In practice, we use the exact same estimator as in Section 3 to estimate the causal version of VOIE. The only difference is the causal interpretability at the cost of additional identifiability assumptions.


\end{document}